\documentclass[%
 reprint,superscriptaddress,
 amsmath,amssymb, nofootinbib,
 aps,twocolumn,pra
]{revtex4-1}
\usepackage{graphicx}
\usepackage{multirow}
\usepackage{amsfonts}
\usepackage{bbold} 
\usepackage{float}
\usepackage{verbatim} 

\usepackage{ulem} 

\usepackage[usenames, dvipsnames]{xcolor}
\usepackage{xifthen}

\newcommand{\eqn}[1]{\begin{eqnarray} #1 \end{eqnarray}}
\newcommand{\tit}[1]{\textit{#1}}

\newcommand{\trm}[1]{\textrm{#1}}
\newcommand{\tsc}[1]{\textsc{\textrm{#1}}}
\newcommand{\tr}[1]{  \textrm{tr}\left[ #1 \right]  }

\newcommand{\zum}[2]{\displaystyle\sum_{#1}^{#2}}

\newcommand{\ket}[1]{| #1 \rangle}
\newcommand{\ketbra}[2]{| #1 \rangle \langle #2 |}

\newcommand{\arxiv}[2][]{\ifthenelse{\isempty{#1}}{\href{http://arxiv.org/abs/#2}{{\tt arXiv:\allowbreak{}#2}}} {\href{http://arxiv.org/abs/#2}{{\tt arXiv:\allowbreak{}#2 [#1]}}}}

\begin{document}

\title{Synthesizing the Born rule with reinforcement learning}

\author{Rodrigo S. Piera\smallskip}
\affiliation{Instituto de Física, Universidade Federal do Rio de Janeiro,
Caixa Postal 68528, Rio de Janeiro, RJ 21941-972, Brazil}
\affiliation{Quantum Research Centre, Technology Innovation Institute, Abu Dhabi, United Arab Emirates}
\author{John B. DeBrota\smallskip}
\affiliation{
Center for Quantum Information and Control, \\
University of New Mexico, Albuquerque, NM 87131, USA}
\author{Matthew B. Weiss\smallskip}
\affiliation{
\href{http://www.physics.umb.edu/Research/QBism/}{QBism Group, University of Massachusetts Boston}, 100 Morrissey Boulevard, Boston, MA 02125, USA}
\author{Gabriela B. Lemos\smallskip}
\affiliation{Instituto de Física, Universidade Federal do Rio de Janeiro,
Caixa Postal 68528, Rio de Janeiro, RJ 21941-972, Brazil}
\author{Jailson Sales Araújo\smallskip}
\affiliation{Instituto de Física, Universidade Federal do Rio de Janeiro,
Caixa Postal 68528, Rio de Janeiro, RJ 21941-972, Brazil}
\author{Gabriel H. Aguilar\smallskip}
\affiliation{Instituto de Física, Universidade Federal do Rio de Janeiro,
Caixa Postal 68528, Rio de Janeiro, RJ 21941-972, Brazil}
\author{Jacques L. Pienaar\smallskip}
\affiliation{
\href{http://www.physics.umb.edu/Research/QBism/}{QBism Group, University of Massachusetts Boston}, 100 Morrissey Boulevard, Boston, MA 02125, USA}





\begin{abstract}
According to the subjective Bayesian interpretation of quantum theory (QBism), quantum mechanics is a tool that an agent would be wise to use when making bets about natural phenomena. In particular, the Born rule is understood to be a decision-making norm, an ideal which one should strive to meet even if usually falling short in practice. What is required for an agent to make decisions that conform to quantum mechanics? Here we investigate how a realistic (hence non-ideal) agent might deviate from the Born rule in its decisions. To do so we simulate a simple agent as a reinforcement-learning algorithm that makes `bets' on the outputs of a symmetric informationally-complete measurement (SIC) and adjusts its decisions in order to maximize its expected return. We quantify how far the algorithm's decision-making behavior departs from the ideal form of the Born rule and investigate the limiting factors. We propose an experimental implementation of the scenario using heralded single photons.
\end{abstract}

\maketitle

\section{Introduction}
One of the most promising contemporary applications for machine-learning is its use in discovering physical laws from raw observational data (for recent surveys see Refs.\;\cite{Dunjko_2018,Roscher2020}). While the most successful examples have tended to exploit prior domain knowledge, more recent work has explored the possibility of `AI scientists': machine-learning models capable of inferring various aspects of physical systems and their dynamics from observed data using only minimal prior assumptions (see e.g.\ Refs.\;\cite{Iten2020,Krenn2022} and citations therein). These models fall naturally within an \tit{agent-environment} paradigm, in which an \tit{agent} (implemented as an algorithm) takes actions in an environment, receives feedback about the consequences of its actions, and thereby adjusts its behavior~\cite{RussellNorvig2009,Brooks1990,Dunjko_2018}.

The \tit{Born rule} in quantum theory determines the probability of observing a measurement outcome given the associated measurement operator and quantum state (see Eq.\,\eqref{eq:Born}). In the literature the rule is either treated as an independent axiom of quantum theory~\cite{chuang00}, or else is logically derived from other axioms; Gleason's celebrated theorem~\cite{Gle} is an example of the latter. Some interpretations of quantum theory, such as QBism~\cite{DeBrota2021} and variants of the many-worlds interpretation~\cite{Deutsch1999,Wallace2009}, argue that the Born rule follows from \tit{decision-theoretic} considerations. 

QBism is particularly well-suited to the agent-environment paradigm because it views quantum theory as a \tit{normative} structure that guides decision-making, rather than a \tit{descriptive} structure that represents reality. A decision-making agent is said to be \tit{rational} to the extent that its probability assignments are coherent\footnote{\tit{Incoherence} for a Bayesian agent is defined as its susceptibility to a \tit{Dutch book}: a series of hypothetical bets, each one acceptable when considered in isolation, but whose net effect is equivalent to a bet that the same agent would never accept.} and its choices maximize its expected utility. An important theoretical result of QBism is that a sufficiently rational agent must either adopt constraints on their probabilities equivalent to the Born rule, or else is provably inconsistent with a set of basic physical assumptions including one motivated by quantum theory~\cite{DeBrota2021}. 

Since QBism regards quantum theory as a normative \tit{ideal}, only agents which are perfectly rational are expected to make decisions that conform precisely to the Born rule, under the relevant assumptions. This naturally raises the question of how much an imperfect agent's betting behaviour might diverge from the precise form of the Born rule, especially when the imperfections reflect realistic constraints on the agent. 

At the most basic level, realistic agents may not be capable of the higher-order reasoning that is assumed of ideal Bayesian agents, nor even capable of reasoning using probabilities that could encode physical assumptions. Even for more advanced agents capable of such reasoning, individual characteristics such as risk-aversion may lead them to prefer actions which do not maximize their expected utility; indeed, human irrationality forms the basis of descriptive modifications of the rational agent paradigm~\cite{Kahneman1979,Tversky1992,Brandstatter2006}. It is thus not at all clear that a given imperfect agent's decision-making behavior would approximate that of an ideal Bayesian agent. 

In the context of machine-learning, the foregoing considerations suggest that even an agent trained on unlimited data obtained by measuring a low-dimensional quantum system under ideal conditions might fall short of perfectly rational behavior. Going further, one could consider the added difficulties presented by finite statistics and experimental error, with an outlook towards studying  increasingly realistic models of agents such as biological systems subject to various kinds of physical and environmental limitations. This would help elucidate the conditions under which an adaptive advantage might exist for agents that make their decisions in accordance with quantum theory.

The agent-environment paradigm is a natural setting to numerically study the behavior of simplified imperfect decision-making agents. Accordingly, in this work we perform a simulation in which a simple reinforcement-learning algorithm is programmed to make `bets' on single particle detection events. The agent adjusts their bets using a standard algorithm and reward function designed to emulate a simple Bayesian agent who is ignorant of quantum theory. We quantify how closely the agent's final betting strategy conforms to the Born rule.

By design, our simplified `proto-Bayesian' agent makes bets which track the long-run relative frequencies of the relevant detection events, exemplifying de Finetti's notion of probabilities as gambling commitments without any self-conscious use of the probability or utility concepts~\cite{Finetti}. Accordingly, we expect that with enough data from a sufficiently rich set of measurements the agent's bets would converge to the Born rule (and indeed our simulation supports this conclusion). We use the simulation to quantify the degree to which the imperfect agent is able to make bets that approximately conform to the Born rule, and we find that its performance is limited mainly by the finite size of the input datasets and the fact that `bets' are chosen from a discrete rather than a continuous set.

We additionally find that the amount of data needed to get reasonably close to the Born rule is significant (requiring the order of $10^5$ simulated measurement outcomes), even in the case of a quantum system with only dimension $d=2$, and assuming ideal experimental conditions. Our findings show that it is quite demanding for an imperfect agent to make decisions that even approximate those of an ideal Bayesian agent who uses quantum theory to guide their decisions; this is true despite the fact that we only require our agent to perform one task, namely to optimize its predictions for the outputs of highly tailored quantum experiment under favorable conditions.

Our work can therefore be considered a first step towards studying more sophisticated models of real agents, such as biological systems constrained by the need to survive in hostile environments. In such cases it is not at all clear whether the payoff to an agent for making decisions that conform to the Born rule would outweigh the costs of adaptively acquiring such refined behavior through experience. Indeed, our preliminary findings seem to indicate that agents subject to more demanding physical, biological or evolutionary constraints would be unlikely to develop decision-making behavior that is sensitive to quantum effects in their environment, except perhaps in highly contrived situations. 

The rest of the paper is organized as follows. In \S\ref{sec:bgtheory}, we provide the relevant QBist background theory for the probabilistic form of the Born rule which motivates our simulations and which allows us to extract a structure from an agent's behavior which we can compare to the Born rule. In \S\ref{sec:algorithm}, we describe our reinforcement learning algorithm, the situations it will face, and its explicit limitations. Then, in \S\ref{sec:simulation}, we expose our agent to simulated data from a two-level system, extract an effective Born rule from its learned behavior, and examine the empirical trend towards convergence to the exact Born rule. In \S\ref{sec:exp}, we propose an optical experiment to test our algorithm. We conclude with a summary and future outlook in \S\ref{sec:summary}.

\section{Background theory}\label{sec:bgtheory}
The Born rule states that the probability of obtaining outcome $j$ when doing measurement $\mathcal{D}$ on state $\rho$ is given by:
\eqn{ \label{eq:Born}
\trm{Pr}(j) = \tr{\rho D_j} \, ,
}
with $\rho$ a density operator and $\mathcal{D} := \{D_j : j=1,2,\dots M \}$ a positive-operator-valued measure (POVM) with mutually exclusive outcomes $j$.

Writing the rule in this form assumes the Hilbert space structure of quantum theory  in terms of which the operators $\rho$, $D_j$ are defined. Since we would like to discover the Born rule from observational data, assuming as little as possible about the mathematical formalism of quantum theory, it will be useful to re-write it in a manner that only refers to probabilities for observable events.

\begin{figure}[!ht]
\includegraphics[width=8cm]{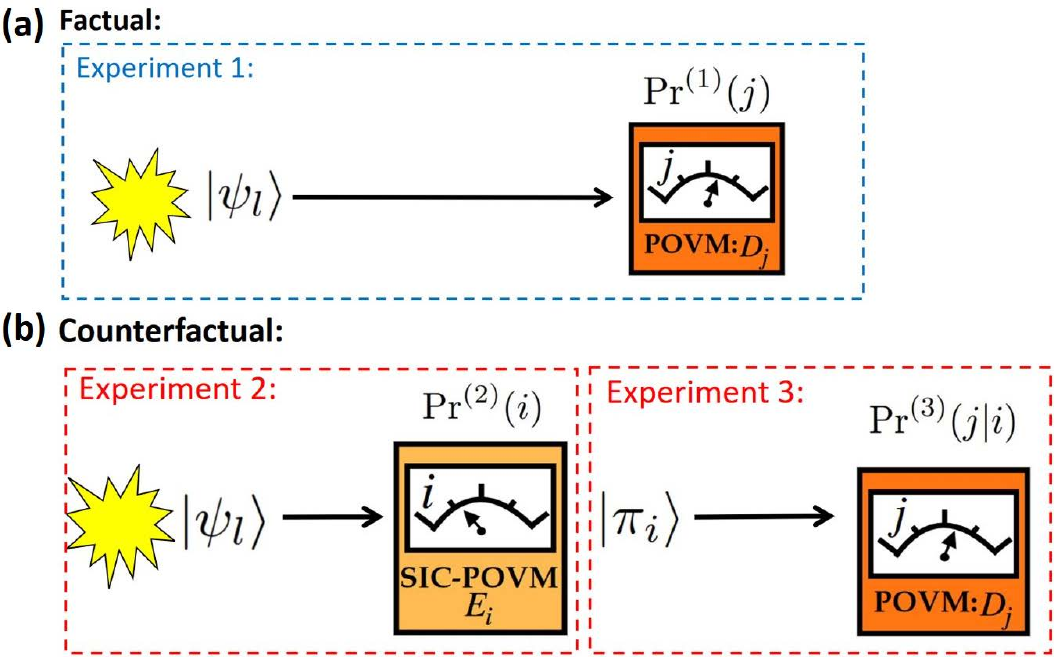}
\caption{(a) The `factual' measurement of POVM $\{ D_j \}$ on state $\ket{\psi_l}$, used to compute $\trm{Pr}^{(1)}(j)$ via the standard Born rule, Eq.\,\eqref{eq:Born}. (b) The `counterfactual' scenario: first $\ket{\psi_l}$ is measured with a SIC-POVM $\{ E_i \}$ (Experiment 2), then state $\ket{\pi_i}$ is prepared according to the outcome and measured by $\{ D_j \}$ (Experiment 3). The probabilities $\trm{Pr}^{(2)}(i)$ and $\trm{Pr}^{(3)}(j|i)$ are used to compute $\trm{Pr}^{(1)}(j)$ using the QBist Born rule, Eq.\,\eqref{eq:Urgleichung}.} 
\label{fig:QBism_scheme}
\end{figure}

To achieve this in the most elegant way possible, QBism conjectures that every Hilbert space of finite dimension $d$ contains a special structure: a set of $d^2$ rank-1 projectors $\Pi_i$ that are equiangular, i.e.
\eqn{
\tr{\Pi_i \Pi_j} = \frac{1}{d+1} \, \quad  \forall i \neq j \, .
}
A set of such projectors, if they exist, defines a symmetric informationally-complete POVM $\{ E_i : i=1,\dots d^2 \}$, where $E_i := \frac{1}{d} \Pi_i$, called a \tit{SIC} (pronounced `seek'). If the conjecture is true, then the Born rule can be uniquely expressed in all finite dimensions as the following relation among probabilities \cite{Fuchs_PR,FuchsStacey2018}:
\eqn{ \label{eq:Urgleichung}
\trm{Pr}^{(1)}(j) = \zum{i=1}{d^2}\, \left( (d+1)\trm{Pr}^{(2)}(i)-\frac{1}{d} \right) \trm{Pr}^{(3)}(j|i) \, ,
}
where 
\eqn{
\trm{Pr}^{(1)}(j) &:=& \tr{\rho D_j} \nonumber \\
\trm{Pr}^{(2)}(i) &:=& \tr{\rho \left( \frac{1}{d} \Pi_i \right)} \nonumber \\
\trm{Pr}^{(3)}(j|i) &:=& \tr{\Pi_i D_j} \, .
}
Conceptually, QBism interprets the probabilities $\trm{Pr}^{(2)}(i)$, $\trm{Pr}^{(3)}(j|i)$, which appear on the right-hand side of \eqref{eq:Urgleichung}, as referring to two  `counterfactual' reference experiments. In the first, the initial system undergoes a SIC-POVM measurement with outcomes $i$; in the second, for each $i$, the system is prepared in the corresponding SIC state $\ket{\pi_i}$ after which it undergoes the $\mathcal{D}$ measurement (see Fig. \ref{fig:QBism_scheme}).
These probabilities are combined to obtain the original probabilities $\trm{Pr}^{(1)}(j)$ using the QBist Born rule, \eqref{eq:Urgleichung}.

Written in this form, we see that the Born rule is equivalent to a formal relation among probabilities for three sets of experimentally observable outcomes, and this relation holds for all states $\rho$ and measurements $\mathcal{D}$. This allows us to generalize the Born rule beyond quantum theory, as we now explain. 

In general, given an informationally complete measurement with outcomes $i$ (not necessarily a SIC, or even a quantum measurement), we can represent any state by its vector of probabilities $\vec{p} := [\trm{Pr}^{(2)}(i) : i=1,\dots,N]$. Similarly any measurement resulting in outcome $j$ can be characterized by its vector of conditional probabilities, $\vec{r}(j) := [\trm{Pr}^{(3)}(j|i) : i=1,\dots,N]$. We can then ask: given a state $\vec{p}$ and measurement $\mathcal{D} := \{\vec{r}(j)  : j=1,2,\dots M \}$, what is the probability $q(j) := \trm{Pr}^{(1)}(j)$ to observe outcome $j$? It can be argued from basic principles that in classical, quantum, and even more general theories, the answer has the form~\cite{DeBrota2021,DFS2020}:    
\eqn{ \label{eq:vecBR}
q(j) = (\vec{r})^{T}(j) \cdot \Phi \cdot \vec{p} \, , 
}
where $\Phi$ is an $N \times N$ real-valued matrix. Specifically, this form follows from (a) assuming reference measurements exist; (b) that the counterfactual probabilities and conditional probabilities are assigned independently of one another; and (c) that measurement outcomes are \textit{noncontextual}. This last condition means that outcomes considered to be the \textit{same} should be assigned the same probability regardless of the fact that they might be consequences of different measurements~\cite{DeBrota2021}. Indeed, this  explains why only the $j$th row of the conditional probability matrix appears in Eq.\,\eqref{eq:vecBR}. In the special case of quantum theory and when the informationally complete measurement is chosen to be a SIC, this equation reduces to the Born rule as written in Eq.\,\eqref{eq:Urgleichung}. We can therefore regard Eq.\,\eqref{eq:vecBR} as a `generalized Born rule' beyond quantum theory.

The specific form of $\Phi$ depends only on the physical theory and the choice of informationally complete measurement used to represent the states and measurements within the theory. For instance, in classical theory there is in principle a measurement that perfectly distinguishes all states of the theory; in that case $\Phi = \mathbb{1}$ (the identity matrix). On the other hand, as is well-known, quantum theory does not allow any measurement that could perfectly distinguish all quantum states, hence $\Phi \neq \mathbb{1}$. It is interesting to ask: if we optimize over all informationally complete quantum measurements, how close to the identity matrix can $\Phi$ get? 

In Ref~\cite{DFS2020} it was proven that the distance to the identity (with respect to any unitarily invariant norm) is minimized in quantum theory when $\Phi$ has $(d+1)-\frac{1}{d}$ on the diagonals and $-\frac{1}{d}$ everywhere else, and this is achieved precisely when the informationally complete measurement is a SIC. We can therefore think of this matrix, $\Phi^{\tsc{SIC}}$, as defining the `quantum Born rule', and assuming we hold the informationally complete measurement fixed, deviations from $\Phi^{\tsc{SIC}}$ can be interpreted as deviations from the Born rule.

In the following, we describe a reinforcement learning algorithm that simulates a simple organism that is totally ignorant of probabilities, but whose actions effectively provide us with values of $q(j)$, $\vec{r}(j)$, and $\vec{p}$ for all $j$. Using a convex optimization procedure subject to the assumptions above, we can extract a $\Phi$ matrix consistent with their behavior and compare it to $\Phi^{\rm SIC}$.

\section{Algorithm and simulations}

\subsection{ Algorithm}\label{sec:algorithm}

Our simulated agent takes as input the measurement outcomes for each of the three experiments in Fig.\,\ref{fig:QBism_scheme} and places different `bets' on which outcomes will occur on the next inputs. By design, our algorithm roughly emulates an ideal Bayesian agent with an exchangeable\footnote{An exchangeable sequence of random variables is one such that any finite permutation of indices preserves the joint distribution. When applied to an infinite sequence, an exchangeable prior is equivalent to a sequence of i.i.d. random variables, conditioned on some distribution~\cite{Bernardo94}.}, full-support, prior distribution who updates to Bayes rule posteriors upon data acquisition.

We divide the learning process into two tiers. In a \textit{learning episode}, the agent bets on the outcomes of a given experiment based on its past experience with this measurement. A learning episode is divided into a large number $\mathcal{S}$ of `steps,' each corresponding to the outcome of a single run of one of the experiments. The learning episodes corresponding to a fixed $\rho$ and $\mathcal{D}$ then make up an \textit{experimental episode}. For instance, for two-level systems with a projective $\mathcal{D}$, an experimental episode comprises 14 different learning episodes: two corresponding to Experiment 1, four to Experiment 2, and eight to Experiment 3 (Fig.\ref{eq:Born}). To distinguish the bets made in different scenarios, we define $B_{s}^{(1)}(j)$, $B_{s}^{(2)}(i)$ and $B_{s}^{(3)}(j|i)$ as the bets made on the experiments 1, 2 and 3, respectively, where the sub-index $s\in [1,\mathcal{S}]$ refers to the step number. The algorithm receives a reward depending only on their choice of bet and on the outcome obtained in each step. For this, we use the quadratic loss function\footnote{A Bayesian rational agent subject to this reward scheme will report their true probabilities in a forecasting scenario~\cite{Bernardo94}. Our algorithm, however, has no explicit awareness of probabilities, which is why we refer to it as a \tit{proto-}Bayesian agent.}, which returns a negative value of $-(B-E)^2$ where $0\leq B\leq 1$ is the amount bet on a given outcome, $E=1$ if that outcome occurs, and $E=0$ otherwise. Depending on this feedback, the agent adjusts their betting behavior in the next step according to a reinforcement-learning (RL) algorithm, and the process is repeated. The number of steps $\mathcal{S}$ is chosen to be large enough so that the agent's preferred choice of bets converge to approximately stable values $B^{(1)}(j)$, $B^{(2)}(i)$, and $B^{(3)}(j|i)$, for a given learning episode, experimental episode and the outcome chosen. We take these convergence betting values to represent the algorithm's effective `posterior probabilities' via the substitutions $B^{(1,2,3)} \leftrightarrow \trm{Pr}^{(1,2,3)}$, which provides us with the components of $q(j),\vec{r}(j),\vec{p}(i)$ for a given experimental episode. These will subsequently be analyzed for conformity to the Born rule. We perform the same process for various experimental episodes, inputting distinct states and performing a variety of measurements $\mathcal{D}_i$. 

In each step $s$, the algorithm `bets' on the detection event by selecting an integer $k \in \{ 0,\dots,N \}$ and defining the bet value as $B^{(1,2,3)}_{s}[k] := \frac{k}{N}$. Since the bets are supposed to stand in for `probability assignments,' the bets should ideally cover \tit{all} values between $0$ and $1$.  However, practically speaking, this would make learning (in the sense we will employ) impossible since the agent would never make the same selection more than once. We therefore restrict the agent to a discrete set of choices, choosing $N$ large enough to minimize the errors that this introduces.

For simplicity, our RL algorithm is based on a standard \tit{epsilon-greedy} algorithm \cite{sutton_RL} concerning the reward function defined above. For example, consider Experiment 2 and suppose the algorithm chooses to bet $k=8$ for outcome $i=3$. If this detector clicks in the $m$-th step, then the algorithm's `reward' for that step is $r_m(B^{(2)}(i=3)[k=8]) = -(\frac{8}{N}-1)^2$, otherwise they receive $r_m(B^{(2)}(i=3)[k=8]) = -(\frac{8}{N})^2$. The agent adds their reward to a running total for each bet, i.e.\ after $\mathcal{S}$ steps the total for bet $r_m(B^{(2)}(i=3)[k=8])$ is $T_{k=8} = \zum{m=1}{M} \, r_m(B^{(2)}(i=3)[k=8])$,  where $M$($\leq \mathcal{S}$) is the number of times the agent chose to bet $k=8$ (omitting the sub-index $s$ for clarity). When deciding which bet to make in each step $s$, the epsilon-greedy algorithm has a $(1-\epsilon)$ chance of choosing the bet with the highest average reward\footnote{If multiple bets share the same running total, it selects randomly between them with equal probability.}  $B^{(2)}(i=3)[\trm{max}_k] :=\{B^{(2)}(i=3)[\trm{k}] : \trm{max}_k(T_k) \}$
and an $\epsilon$ chance of choosing any bet uniformly at random. This algorithm thereby represents a balance between `exploitation' (choosing bets that have been beneficial in the past) and `exploration' (taking a risk by trying other bets).


At the end of the learning episode, the algorithm determines the best bet price based on the maximum average reward accumulated over the episode. As such, our agent mimics a de Finetti style gambler (albeit without a self-conscious notion of coherence) whose bets we can interpret as stand-ins for probability assignments. Moreover, a result of de Finetti tells us that the posterior expectation for a truly Bayesian agent starting with an exchangeable, full support prior distribution will track relative outcome frequencies in the infinite data limit. Thus, there is a tight relationship between the three concepts of the bet, the probability, and the relative frequency.

The algorithm proceeds analogously for different experimental episodes, outputs, and bets, obtaining all the probabilities involved in $q(j),\vec{r}(j),\vec{p}(i)$.  The $\Phi$  matrix is then reconstructed by minimizing  the Frobenius norm, defined as $\big( q(j) - (\vec{r})^{T}(j) \cdot \Phi \cdot \vec{p} \big)^2 $, over choices of $\Phi$ for all  $q(j),\vec{r}(j),\vec{p}$.

\subsection{Simulation }\label{sec:simulation}

\begin{figure*}[!th]
\includegraphics[width=16cm]{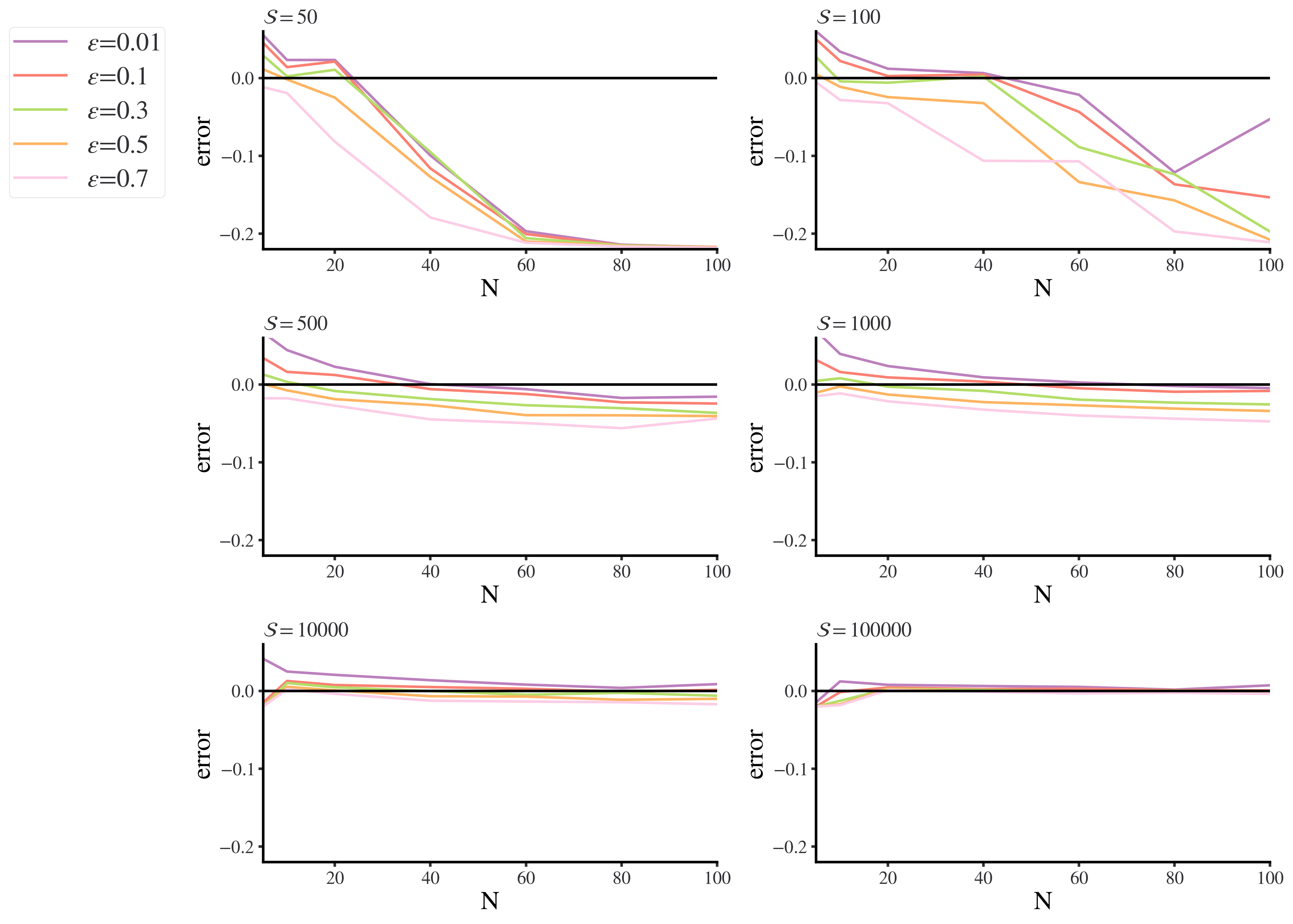}
\caption{ 
On what basis should we choose the parameters for our agent? To explore this question, we generate 50 datasets of binary values generated randomly with $p=0.22$, and consider running our agent for different numbers of steps $(\mathcal{S})$, with different values for the bet quantization $(N)$, and for different values of the parameter $\epsilon$. For each tuple of parameters $(\mathcal{S}, N, \epsilon)$, we train 50 agents on the same dataset, and take the average of the final optimal bet value. Finally,  we repeat the process for each of the 50 datasets, averaging over the results. The consequence is depicted above: each of the six plots represents a different choice of $\mathcal{S}$; the colors of the plotted lines denote a choice of $\epsilon$, and we plot the error $\langle \hat{p}\rangle - 0.22$ as a function of $N$. We can see that as long as $N$ is above a certain threshold, the choice of $\epsilon$ is largely irrelevant as $\mathcal{S}$ grows large.
} 
\label{fig:sweep}
\end{figure*}

Before studying the performance of our algorithm with simulated datasets, we investigate which parameter region best minimizes the errors of convergence. For this, we set the simulated probability $p=0.22$ and  plot the error, defined as $\langle\hat{p}\rangle-0.22$, as a function of $N$ for different values of $\mathcal{S}$ and $\epsilon$. The results are depicted  in Fig.\,\ref{fig:sweep}, showing that above $\mathcal{S}=10^4$, the behaviour of the algorithm is independent of the choice of $\epsilon$. Thus, from now on we work with $\mathcal{S} >10^4$, $N=50$, and we arbitrarily choose $\epsilon=0.5$.

\begin{figure}[!ht]
\includegraphics[width=9cm]{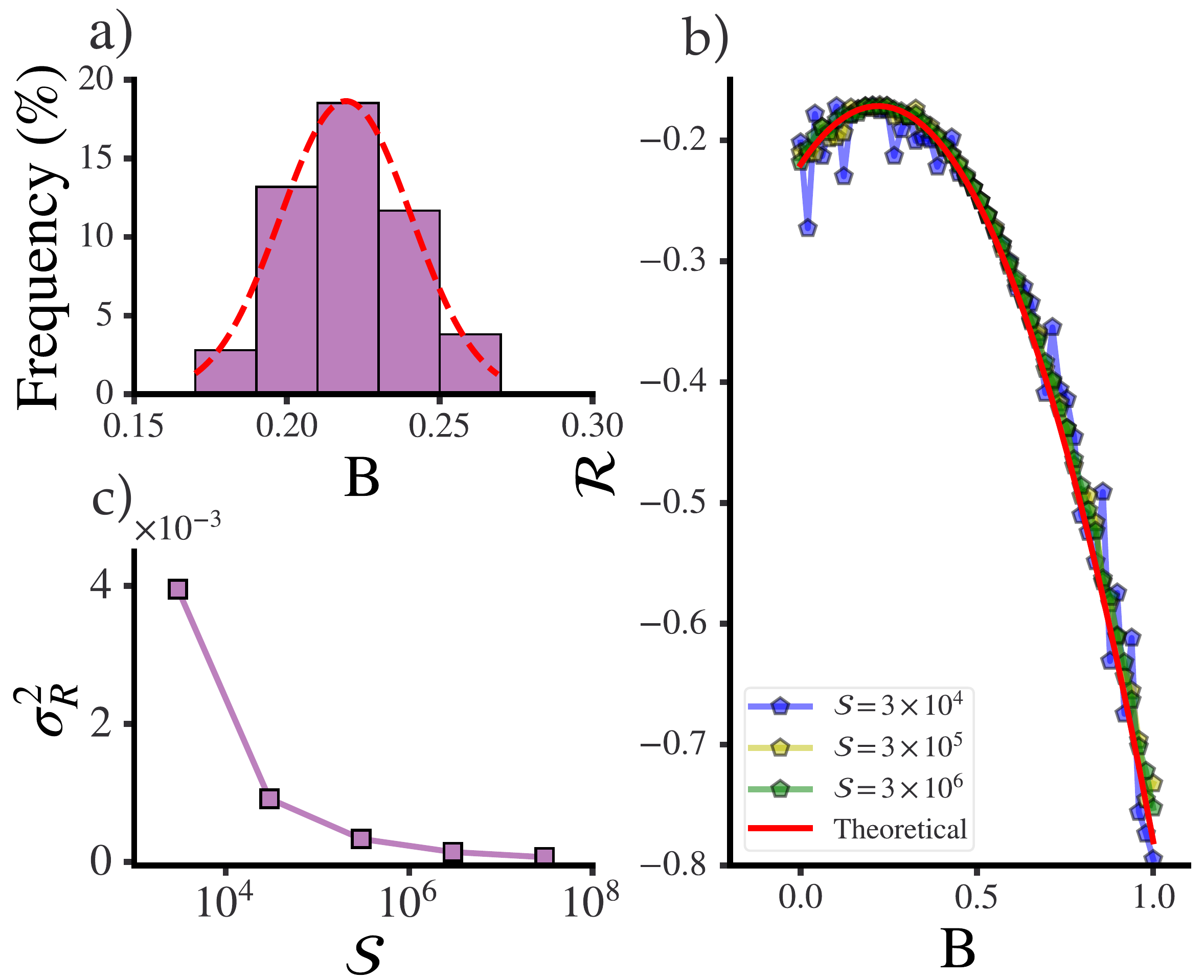}
\caption{ 
(a) Relative frequencies of the best actions chosen by our agent as a function of the bet $B$ on a given outcome. The results are for one learning episode, in which we fed the algorithm with $200$ different simulated datasets created with probability $p=0.22$. The red dashed line is a Gaussian fit for the histogram and is centered on the `true' value of the probability. (b) The calculated expected total reward $\mathcal{R(B)}$ for $p=0.22$ (Eq.\,\ref{eq:reward}).
 Our agent calculates  $\mathcal{R}$ over $\mathcal{S}$ steps and returns the action that gives the highest reward (the maximum point in the graph). The red line shows the theoretical curve for the case of $p=0.22$. The purple, yellow and green marks/lines represent the values calculated by our agent for $\mathcal{S}=3\times10^4,\; 3\times10^5, \;3\times10^6 $, respectively.   (c) Convergence of our algorithm to a set of 200 randomly chosen probabilities $p_i$.
} 
\label{fig:example_runs}
\end{figure}

To better understand the performance of our RL algorithm's ability to converge to a particular probability, consider Fig.\,\ref{fig:example_runs}(a). We plot the average resulting betting frequencies obtained by our algorithm after 200 runs of $\mathcal{S}=3\times 10^5$ steps. One can see that the largest frequency is obtained at $B=0.22$, showing that the algorithm mostly converges to $p$, although there is some variance due to the finite statistics of the input data set. By fitting the histogram in Fig.\,\ref{fig:example_runs}(a) with a Gaussian distribution (red dashed line), we observe that its center coincides with $p$ and its standard deviation is $\sigma_E=0.02$.
  
In Fig.\,\ref{fig:example_runs}(b) we plot the expected reward as a function of $B$ for $p=0.22$, which can be computed as 
\begin{equation}\label{eq:reward}
    \mathcal{R}=p(1-B)^2 + (1-p)B^2, 
\end{equation}where $p$ and ($1-p$) are the relative frequency of a certain outcome occurring ($E=1$) or not occurring ($E=0$), respectively.
For a finite number of steps  $\mathcal{S}$, the function $\mathcal{R}$ is susceptible to fluctuations, which cause our results to deviate from the theoretical expectations (red line in Fig.\,\ref{fig:example_runs}(b)).  Furthermore, the smaller $\mathcal{S}$, the greater the fluctuations. This variability results in peaks on the plots, which are comparable to the maximum value of the theoretical line but occur at a different $B_{k}$ value. Consequently, this leads to convergence to a $p$ value different from the theoretical curve's maximum.

We expand our analysis to study the convergence of our algorithm to a set of 200 randomly chosen probabilities, each denoted as $p_{i}$. We generate datasets with variable size $\mathcal{S}$. Each list is then fed into our algorithm, and we compute the square difference $\sigma_R^{2}=\sum_{i}^{200}(B_{i}-p_{i})^2$ between the eventual bet price and the corresponding probability. The results shown in Fig.\,\ref{fig:example_runs}(c) demonstrate a steady reduction in this difference as $\mathcal{S}$ increases. This trend was observed regardless of the specific learned probability value.  
Hence, the precise convergence of our algorithm consistently occurs across a broader spectrum of probabilities, demonstrating its reliability and robustness.

\begin{figure}[h]
\includegraphics[width=8cm]{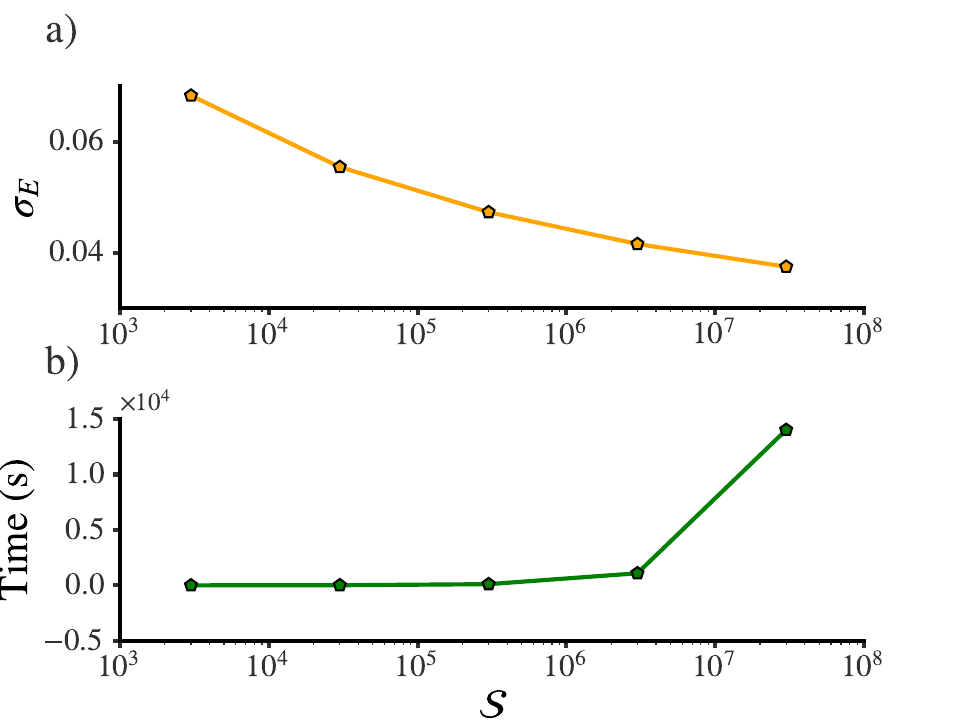}
\caption{Precision of algorithm convergence. (a) The standard deviation of the probability distribution of our agent around the `true' value of the probability ($p=0.23$ in this case) as a function of the number of steps. It is possible to increase the precision of our agent by increasing the number of steps. (b) The average time it takes for our agent to learn the probability as a function of the number of steps. 
} 
\label{fig:convergence}
\end{figure}

Finally, in Fig.\,\ref{fig:convergence}, we can see that $\sigma_E$ monotonically decreases as $\mathcal{S}$ increases, at the cost of a longer average time of simulation. Based on these explorations we choose $\mathcal{S}=3\times 10^5$ steps for each learning episode for the simulated experiments of interest.

When we apply our algorithm to a two-level system, for example, the polarization state of single photons,  the $\Phi$ matrix in Eq.\,\eqref{eq:vecBR} is a $4\times4$ matrix. In this case, for a particular experimental episode, the input state of Experiments 1 and 2 is some qubit state and, for Experiment 3, the input states range over the SIC-states $\{\ketbra{\pi_i}{\pi_i}\}_{i=1}^4$. The detection events for Experiment 2 are always the four outcomes of the SIC-POVM $\{E_i\}_{i=1}^{4}$, while for Experiments 1 and 3, the detection events are the outcomes of a chosen POVM $\{D_j\}$.

In order to estimate the $\Phi$ matrix for simulated data, we construct data sets with known probabilities $\trm{Pr}^{(1-3)}$ compatible with the scenario in Fig.\,\ref{fig:QBism_scheme}, such that the relation of these probabilities accords with the Born rule, Eq.\,\eqref{eq:Urgleichung}, and train our agent on them. Then, once we obtain the quantities $q(j)$, $\vec{r}(j)$, $ \vec{p}$, we minimize the Frobenius norm of the difference between the left and the right-hand side of Eq.\,\eqref{eq:vecBR} over the chosen experimental episodes to arrive at a synthetic $\Phi$  matrix summarizing the learned behavior of the algorithm.

\begin{figure}[!ht]
\includegraphics[width=9cm]{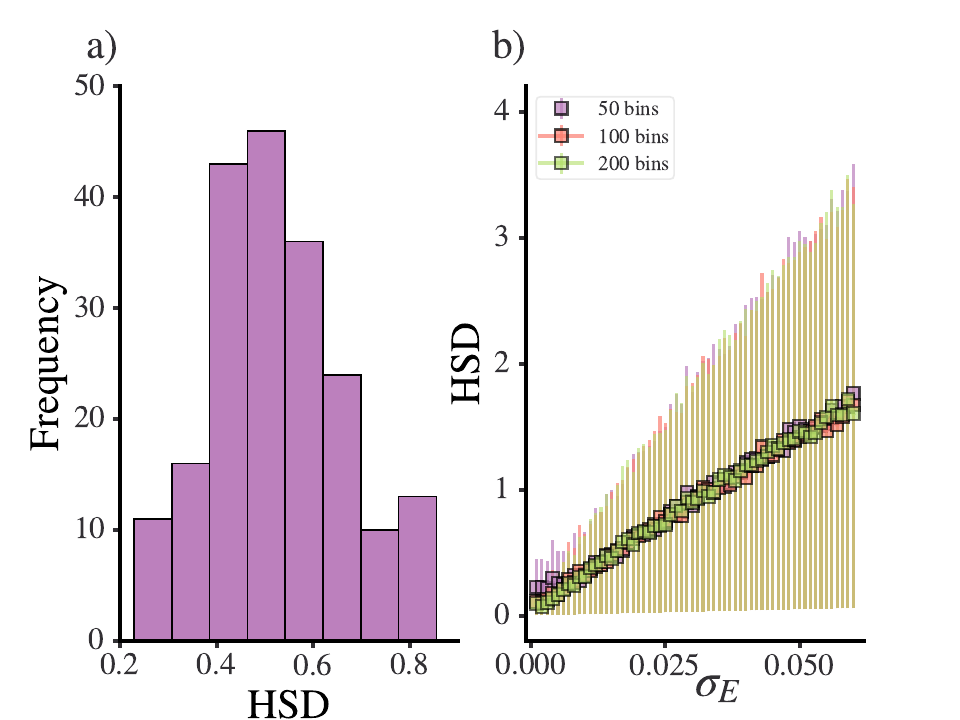}
\caption{(a) Histogram of the HSD values between the learned $\Phi$ matrix and $\Phi^{\rm SIC}$ for 200 runs of our simulation, each with $\mathcal{S}=3\times 10^5$. (b) HSD values obtained from sampling discretized Gaussian distributions in place of the best bets we would normally obtain from the algorithm in order to address how the convergence of the learned $\Phi$ matrix depends on the variability of the best bet. We consider three values of $N$ and report the standard deviation $\sigma_\Phi$ around each mean.} 
\label{fig:trace_simulation}
\end{figure}

We implement this procedure for $90$ different experimental episodes corresponding to $30$ randomly chosen pure input states and measurements of the $3$ Pauli operators. The results are shown in Fig.\,\ref{fig:trace_simulation}(a), where we calculate the Hilbert-Schmidt distance (HSD)  between the synthesized $\Phi$ matrix and the theoretical matrix given by  
\begin{equation}
\Phi^{\rm SIC}=
    \begin{pmatrix}
        5/2 & -1/2 & -1/2 & -1/2\\
        -1/2 & 5/2 & -1/2 &-1/2 \\
        -1/2 & -1/2 & 5/2 &-1/2\\
        -1/2 & -1/2& -1/2 &  5/2
    \end{pmatrix}.
\end{equation}
The mean value is $\mu_\Phi=0.52$ with standard deviation $\sigma_\Phi=0.14$ for 200 runs. For reference, the HSD of $\Phi^{\text{SIC}}$ to the identity is $2\sqrt{3} \approx 3.46$ and the HSD to the `garbage' $\Phi$ whose entries are all $1/4$ (so that the columns sum to one, furnishing a probabilistic model) is $3\sqrt{3} \approx 5.2$. This means that most of the obtained $\Phi$ matrices are comparatively close to the theoretical matrix, although deviation from the null value remains significant. 

From this analysis, some questions naturally arise. For instance, how close can $\mu_\Phi$ be to zero? How can we reduce $\sigma_\Phi$? To address the potential ideal convergence of the HSD, instead of using the probabilities obtained from our algorithm, we consider probabilities sampled from discretized Gaussian probability distributions centered at the expected probabilities. For a set of probabilities compatible with the scenario in Fig.\,\ref{fig:QBism_scheme}, we study how the HSD behaves as a function of the standard deviation $\sigma_E$ of these distributions (we suppose here that all the involved distributions have the same tunable $\sigma_E$). The outcomes are shown in Fig.\,\ref{fig:trace_simulation}(b).
We note that the mean values of the HSD distributions, resembling those in Fig.\,\ref{fig:trace_simulation}(a), approach zero as $\sigma_E$ is reduced. Moreover, the size of $\sigma_\Phi$, indicated by bars, also decreases with smaller values of $\sigma_E$.
Based on these findings, we conclude that as $\sigma_E$ shrinks, the HSD distribution becomes more concentrated at zero. 

Our results indicate that in order to achieve meaningful convergence to the Born rule, our agent requires substantial resources both in terms of data and time, which a more realistic agent may not have at their disposal due to survival pressures. We expect that these difficulties would increase when the agent is faced with the limited quantity and noise typical of experimental data. We propose a possible experimental realization in the following section.

\section{Experimental Proposal}\label{sec:exp}

\begin{figure}
\includegraphics[width=8cm]{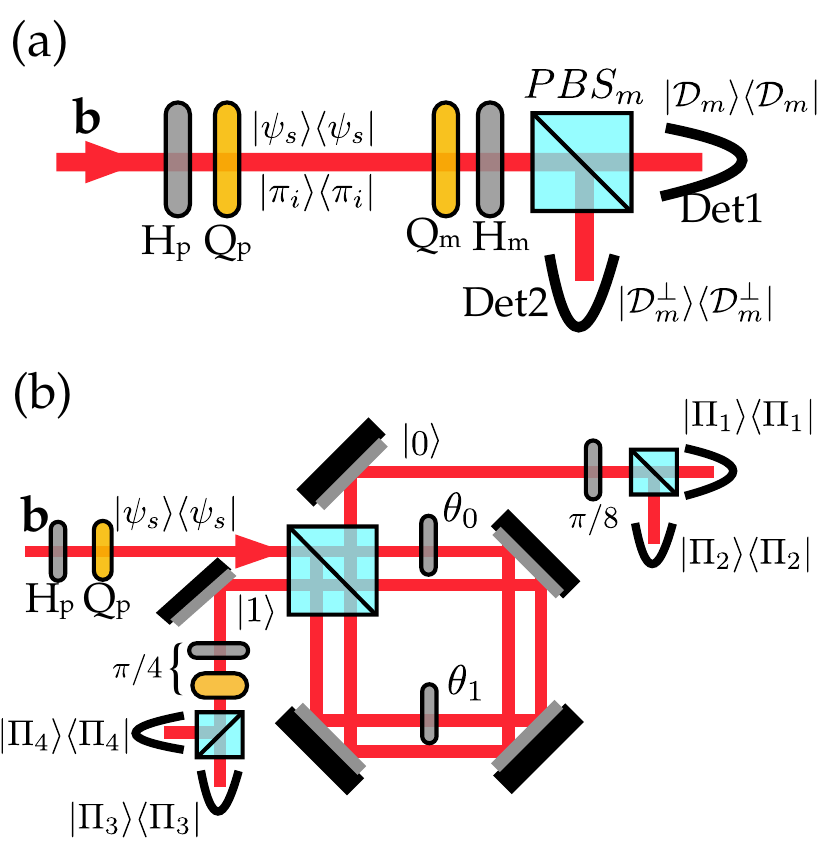}
\caption{The proposed experimental setup. (a) $B^{(1)}(j)$ and $B^{(2)}(j|i)$ are learned with the data obtained from the detections in Det1 or Det2  after preparing the polarization of the photons in the states $\ketbra{\psi_s}{\psi_s}$ and $\Pi_i$, respectively, using H$_p$ and Q$_p$ and projective measurements implemented by Q$_m$, H$_m$ and PBS$_m$. (b) $B^{(2)}(i)$ learning requires a SIC-POVM measurement, which is performed using the displaced Sagnac interferometer.} 
\label{fig:Setup}
\end{figure}

Here we propose an optical experiment to test our algorithm with real data. For this, we suggest using the polarization states of heralded single photons from spontaneous parametric-down conversion \cite{kwiat99}. The signal photons are injected in mode $b$ of Fig.\,\ref{fig:Setup} while the idler photons (not shown) are used to herald detection events. The photons in $b$ are prepared in quasi-pure polarization states, using the wave-plates H$_p$ and Q$_p$. In the paraxial approximation, the polarization of the signal photons is a two-level system. 
 
The two experimental set-ups depicted in Fig.\,\ref{fig:Setup} are used to implement Experiments 1-3, shown in Fig.\,\ref{fig:QBism_scheme}. In this scheme, Experiment 1 is implemented according to Fig.\,\ref{fig:Setup}(a), preparing some set of states $\{\ketbra{\psi_s}{\psi_s}\}$ and measuring them using an appropriate POVM $\mathcal{D}$. See Appendix\,\ref{app:nonclassical} for a discussion of how to choose these states and measurements to ensure nonclassical statistics. The POVM outcomes $j$ are then fed into the algorithm and this is repeated until a stable value of $B^{(1)}(j)$ is obtained. 
Experiment 2 is implemented according to Fig.\,\ref{fig:Setup}(b). The polarization states $\ketbra{\psi_s}{\psi_s}$ are now measured using the four-outcome SIC-POVM $\{ \frac{1}{2} \Pi_i \}$ \cite{Kurtsiefer06, Gomes16}. 
 
To implement the SIC-POVM, photons are directed through a displaced Sagnac interferometer with four output modes, shown in Fig.\,\ref{fig:Setup}(b), such that detection of a photon in the $i$-th output corresponds to the $i$-th outcome of the SIC-POVM. The displaced Sagnac interferometer implements a unitary operation $U_{\text{S}}$ described by the  following map 
\begin{eqnarray}
\ket{H}\ket{0}&\rightarrow&x\ket{H}\ket{0}-y\ket{V}\ket{1}\nonumber\\
\ket{V}\ket{0}&\rightarrow&y\ket{V}\ket{0}+x\ket{H}\ket{1},
\label{eq:transfPPBS}
\end{eqnarray}
 where  $x$ and $y$ are related with the half waveplates inside the interferometer, such that 
 \begin{equation}
 \begin{split}
     x&=\text{cos}(\theta_{0}/2)=\text{sin}(\theta_{1}/2)\\
     y&=-\text{cos}(\theta_{1}/2)=\text{sin}(\theta_{0}/2)\;,
     \end{split}
 \end{equation}
where $\theta_{0}$ and $\theta_{1}$ are the angles of the fast-axis of the half waveplates with respect to the horizontal polarization.  The spatial modes $\ket{0}$ and $\ket{1}$ correspond to the spatial modes of the interferometer. To compensate for any extra phase appearing in the interferometer, two tilted quarter waveplates are added at its outputs.  A half waveplate followed by a quarter waveplate, both at an angle $\pi/4$, are inserted in mode $\ket{1}$, while a half waveplate at $\pi/8$ is inserted in mode $\ket{0}$. The net transformation of these plates is described by a unitary operation on the polarization degree of freedom $U_{\text{WPS}}=U_{\text{HWP}}^{1}(\pi/4) \cdot U_{\text{QWP}}^{1}(\pi/4)\cdot U_{\text{HWP}}^{0}(\pi/8)$, where 
 \begin{equation}
\text{U}_{\text{QWP}}(\varphi)=\frac{1}{\sqrt{2}}
\begin{pmatrix}
i-\text{cos}(2\varphi) & \text{sin}(2\varphi) \\
\text{sin}(2\varphi) & i+\text{cos}(2\varphi)
\end{pmatrix},
\label{eq:QWP}
\end{equation}
\begin{equation}
\text{U}_{\text{HWP}}(\varphi)=
\begin{pmatrix}
\text{cos}(2\varphi) & -\text{sin}(2\varphi) \\
-\text{sin}(2\varphi) & -\text{cos}(2\varphi)
\end{pmatrix}.
\label{eq:HWP}
\end{equation} 
 After PBS$_A$ and PBS$_B$, the complete unitary transformation can be calculated as $U_{\text{SA}}=U_{\text{PBSs}}\cdot U_{\text{WPS}}\cdot U_{\text{S}}$. By tracing out the path degree of freedom, we  calculate the POVM elements:
\begin{equation}
\Pi_1=
\frac{1}{2}\begin{pmatrix}
x^{2} & -x y  \\
-xy  & y^{2} 
\end{pmatrix},
\label{eq:POVM1}
\end{equation}
\begin{equation}
\Pi_2=
\frac{1}{2}\begin{pmatrix}
x^{2} & xy  \\
xy  & y^{2} 
\end{pmatrix},
\label{eq:POVM2}
\end{equation}
\begin{equation}
\Pi_3=
\frac{1}{2}\begin{pmatrix}
y^{2} & -ixy \\  
ixy  & x^{2}  
\end{pmatrix},
\label{eq:POVM3}
\end{equation}
\begin{equation}
\Pi_4=
\frac{1}{2}\begin{pmatrix}
y^{2} & ixy  \\ 
-ixy  & x^{2}  
\end{pmatrix},
\label{eq:POVM4}
\end{equation}
where after setting $x^2=\frac{1}{2} + \frac{1}{2\sqrt{3}}$ and $y^2=\frac{1}{2} - \frac{1}{2\sqrt{3}}$, a SIC-POVM is obtained \cite{PhysRevA.74.022309}. Feeding data from this experiment into the algorithm produces $B^{(2)}(i)$.

Finally, Experiment 3 is implemented according to Fig.\,\ref{fig:Setup}(a), except now the input photons are prepared in each of the four SIC states $\Pi_i$. By feeding this data into the algorithm one obtains $B^{(2)}(j|i)$.

\section{Summary and Outlook}\label{sec:summary}

We have studied how closely a simple reward-seeking algorithm can approximate decision-making behavior consistent with the quantum mechanical Born rule. To do this, we take advantage of a theorem due to QBism that allows us to formulate the Born rule as a decision-theoretic constraint on an agent's betting strategy. We evaluated our algorithm on simulated data and also proposed a photonic experiment to produce ``real'' data.

Our work is intended to be the first step towards more detailed investigations of how realistic and imperfect decision-making agents could deviate from optimal behavior in a quantum-theoretic context. It raises the fundamental question of whether biological constraints and the urgency of survival in realistic settings might inhibit agents' sensitivity to quantum effects.

The scenario we have investigated here is limited in four essential ways, opening up several avenues for generalizations in future work, which we now discuss.

First, our algorithm is trained on simulated data in which quantum effects are, in principle, straightforward to discern at the statistical level. A more realistic agent, even supposing it is highly sensitive to quantum fluctuations in its environment, would also require the input of additional data relevant to its survival, which would somewhat obfuscate the quantum signal. 

Second, by using the quadratic loss function as our reward function, we have effectively guaranteed that the agent's bets will track the long-run relative frequencies of the detection events so that its behavior is expected to converge to that of an ideal Bayesian agent given enough input data. However, realistic systems (such as humans), even ones capable of using probabilities to compute the decision that would maximize their expected utility, are also subject to various complex biological and survival pressures that often lead them to behave in ways that are sub-optimal from a Bayesian point of view~\cite{Allais1953,Ellsberg1961}. Future work should therefore consider how best to model realistic but simple non-Bayesian agents. 

Third, our agent algorithm itself does not output the Born rule (or any approximation to it); rather, it merely takes actions that \tit{we} can summarize analytically using the reconstructed $\Phi$ matrix, which then provides a synthetic overview of its behavior than can be easily compared to the $\Phi$ matrix that represents the Born rule via the QBist Eq.\,\eqref{eq:vecBR}. The form of this equation is based on an assumption of noncontextuality which could be relaxed in future work.

Finally, our reinforcement-learning algorithm represents an extremely simple agent that does not explicitly assign probabilities or make inferences. By contrast, realistic models of biological systems are typically \tit{multiscale systems} wherein complex decision-making behavior at the macro-level can be shown to emerge from simple ``mindless" reward-seeking behavior at the micro-level; a paradigmatic example is the modeling of ant colonies as agents that make active inferences, even though the individual ants follow extremely simple behavior patterns that merely react to pheromone traces left by other ants~\cite{friedman_ant_colony_2021}. It would be interesting to investigate whether simple reinforcement learning agents like the one studied here could similarly serve as building blocks for more complex forms of emergent agency and whether the multiscale optimization of such systems could lead to quantum-sensitive behavior.

We conclude by noting a possible direction for future work. Cognitive scientist Donald Hoffman and collaborators have presented results which suggest that simulated organisms which are sensitive to all information present in their environment are ruthlessly selected against in favor of those that are only sensitive to information directly relevant to their survival~\cite{mark_natural_2010}. In our work, we essentially impose a fitness function which is tuned precisely to the ``truth'' represented by the detector outcomes, and we find that this already places a heavy burden on our simple agent. While it is certainly possible that an agent with an alternative structure, indeed, one tailored by evolution, might be more adept at internalizing quantum effects in the way we have considered, it is also possible that any situation where survival is decoupled from truth would militate against agents ``learning the Born rule.'' It would thus be fruitful to conduct more sophisticated quantum game-theoretical simulation in hopes of understanding those conditions under which quantum-aware life might arise.

\acknowledgments
This publication was made possible through the support of Grant 62424 from the John Templeton Foundation. The opinions expressed in this publication are those of the authors and do not necessarily reflect the views of the John Templeton Foundation. Matthew B. Weiss was further supported by National Science Foundation Grant 2210495, and John B. DeBrota acknowledges support from National Science Foundation Grant 2116246. This work was also supported in part by the John E.\ Fetzer Memorial Trust. The authors acknowledge financial support from the Brazilian agencies CAPES and CNPq (PQ Grants No. 305740/2016-4, 307058/2017-4, 309020/2020-4 and INCT-IQ 246569/2014-0). GHA acknowledges FAPERJ (JCNE E-26/201.355/2021) and FAPESP (Grant No. 2021/96774-4). GBL also acknowledges FAPERJ (JCNE E-26/201.438/2021).

\normalem
\bibliography{references.bib}

\appendix*

\renewcommand{\thesubsection}{\Alph{subsection}}



\subsection{Certification of Nonclassicality}\label{app:nonclassical}

\begin{figure}[H]
\includegraphics[width=\linewidth]{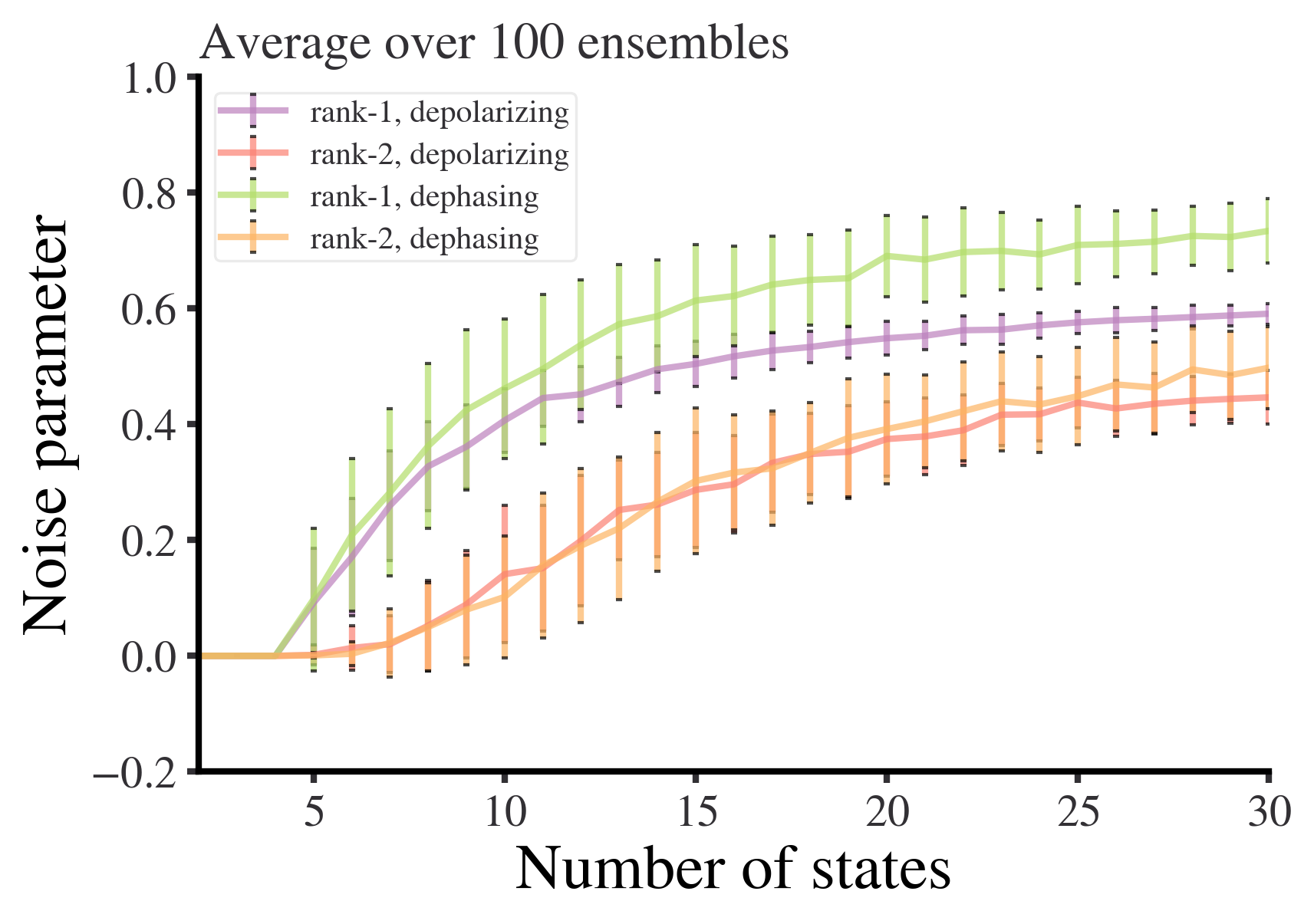}
\caption{ 
Here we plot the minimum noise parameter necessary for the existence of a noncontextual ontological model as a function of the number of (randomly sampled) states. We compare both rank-1 and rank-2 states, as well as two types of noise---depolarizing noise and dephasing noise---averaging over 100 ensembles in each case.
} 
\label{fig:nchv}
\end{figure}

Reference \cite{selby_accessible_2023} provides a scheme for determining whether a fragment of quantum theory has a classical explanation in the sense of a \emph{noncontextual ontological model}.  The provided algorithm calculates whether there is a linear embedding of some set of states into the probability simplex and simultaneously a linear embedding of some set of POVM elements (effects) into the hypercube dual to the probability simplex, such that all probabilities are preserved. Such a \emph{simplex embedding} corresponds directly to a noncontextual ontological model. In the presence of sufficient noise, however, any fragment has a classical explanation in this sense. Thus one can consider a relaxation of the problem where one quantifies the nonclassicality of a fragment by the amount of noise (according to some noise model) that must be added to the states (or equivalently the effects) for them to be simplex embeddable. 

In particular, if we encode the effects as the rows of a matrix $\textbf{E}$ and the states as the columns of a matrix $\textbf{S}$, we require that
\begin{align}
    P(E|S) = \textbf{ES}= (\textbf{E}\sigma_E) (\sigma_S \textbf{S}) = P(E|\lambda)P(\lambda|S),
\end{align}
where $\sigma_E, \sigma_S$ perform the embeddings, enabling one to think of a set of ``hidden variables'' $\lambda$, and where we have expressed everything succinctly in matrix notation. So $\sigma_E\sigma_S$ ought to act as the identity on the relevant space; relaxing the problem amounts to allowing $\sigma_E\sigma_S$ to be instead, for example,  a depolarizing channel $\mathcal{E}_p(\rho) = p I/d + (1-p)\rho$ or a $Z$-dephasing channel $  \mathcal{E}_p(\rho) =  \left(1- \frac{1}{2}p\right)\rho +  \frac{1}{2}p \hat{Z} \rho \hat{Z}$.

If we take the set of states to be the Pauli eigenstates and the SIC projectors, and the set of effects to be Pauli projectors and the SIC-POVM elements, then to find a noncontextual ontological model, we require
\begin{align}
    p_{\text{depolarizing}} \approx 0.413 &&  p_{\text{dephasing}} \approx 0.366.
\end{align}
More generally, keeping the effects fixed to the SIC-POVM and the three Pauli measurements, we can explore how increasing the number of randomly sampled states makes finding a noncontextual ontological model increasingly difficult (Fig.\,\ref{fig:nchv}).

\end{document}